\documentclass[11pt,english]{article}
\usepackage{amsfonts}
\usepackage{amsmath}
\usepackage{geometry}
\usepackage{amsthm}
\usepackage{graphicx}
\usepackage{mathrsfs}
\usepackage{enumerate}
\usepackage{color}
\usepackage{etex,etoolbox}
\usepackage[T1]{fontenc}
\usepackage[latin9]{inputenc}

\makeatletter
\providecommand{\@fourthoffour}[4]{#4}
\def\fixstatement#1{%
  \AtEndEnvironment{#1}{%
    \xdef\pat@label{\expandafter\expandafter\expandafter
      \@fourthoffour\csname#1\endcsname\space\@currentlabel}}}

\globtoksblk\prooftoks{1000}
\newcounter{proofcount}

\long\def\proofatend#1\endproofatend{%
  \edef\next{\noexpand\begin{proof}[Proof of \pat@label]}%
  \toks\numexpr\prooftoks+\value{proofcount}\relax=\expandafter{\next#1\end{proof}}
  \stepcounter{proofcount}}

\def\printproofs{%
  \count@=\z@
  \loop
    \the\toks\numexpr\prooftoks+\count@\relax
    \ifnum\count@<\value{proofcount}%
    \advance\count@\@ne
  \repeat}
\makeatother

\DeclareMathOperator*{\argmax}{arg-max}

\newtheorem{theorem}{Theorem}[section]
\newtheorem{thm}{Theorem}[section]

\newtheorem{lem}[theorem]{Lemma}
\newtheorem{proposition}[theorem]{Proposition}
\newtheorem{corollary}[theorem]{Corollary}

\newtheorem{example}{Example}[section]

\fixstatement{thm}
\fixstatement{lem}

\makeatother

\usepackage{babel}

\newcommand{\musiclab}{\textsc{MusicLab}}

\usepackage{xcolor}

\begin{document}

\title{On the Optimality and Predictability \\ of Cultural Markets with Social Influence}
\author{
P. Van Hentenryck\footnote{National ICT Australia (NICTA)  and The Australian National University (pvh@nicta.com.au)}
\and
A. Abeliuk\footnote{National ICT Australia (NICTA) and The University of Melbourne (Andres.Abeliuk@nicta.com.au)}
\and
F. Berbeglia\footnote{National ICT Australia (NICTA) (Franco.Berbeglia@nicta.com.au)}
\and
G. Berbeglia\footnote{National ICT Australia (NICTA) and the Centre for Business Analytics, Melbourne Business School (g.berbeglia@mbs.edu)}
}

\maketitle

\begin{abstract}
  Social influence is ubiquitous in cultural markets, from book
  recommendations in Amazon, to song popularities in iTunes and the
  ranking of newspaper articles in the online edition of the New York
  Times to mention only a few. Yet social influence is often presented
  in a bad light, often because it supposedly increases market
  unpredictability.

  Here we study a model of trial-offer markets, in which participants
  try products and later decide whether to purchase. We consider a
  simple policy which ranks the products by quality when presenting
  them to market participants. We show that, in this setting, market
  efficiency always benefits from social influence. Moreover, we prove
  that the market converges almost surely to a monopoly for the
  product of highest quality, making the market both predictable and
  asymptotically optimal.  Computational experiments confirm that the
  quality ranking policy identifies ``blockbusters'' in reasonable
  time, outperforms other policies, and is highly predictable. 

  These results indicate that social influence does not necessarily
  increase market unpredictability. The outcome really depends on how
  social influence is used.
\end{abstract}

\section{Introduction}

Social influence is ubiquitous in cultural markets. From book
recommendations in Amazon, to song popularities in iTunes, and article
rankings in the online version of the New York Times, social influence
has become a critical aspect of the customer experience.  Yet there is
considerable debate about its benefits and its effects on the market.

In their seminal paper~\cite{salganik2006experimental}, Salganik,
Dodds, and Watts argued that social influence makes markets more
unpredictable, providing an explanation about why prediction in
cultural markets is a wicked problem. To investigate this hypothesis
experimentally, they created an artificial music market called the
\musiclab{}. Participants in the \musiclab{} were presented a list of
unknown songs from unknown bands, each song being described by its
name and band. The participants were divided into two groups exposed
to two different experimental conditions: the {\em independent}
condition and the {\em social influence} condition. In the first group
(independent condition), participants were provided with no additional
information about the songs. Each participant would decide which song
to listen to from a random list. After listening to a song, the
participant had the opportunity to download it. In the second group
(social influence condition), each participant was provided with an
additional information: The number of times the song was downloaded by
earlier participants. Moreover, these participants were presented with
a list ordered by the number of downloads. Additionally, to
investigate the impact of social influence, participants in the second
group were distributed in eight ``worlds'' evolving completely
independently. In particular, participants in one world had no
visibility about the downloads and the rankings in the other worlds.

The \musiclab{} is a trial-offer market that provides an experimental
testbed for measuring the unpredictability of cultural markets. By
observing the evolution of different worlds given the same initial
conditions, the \musiclab{} provides unique insights on the impact of
social influence and the resulting unpredictability. In particular,
Salganik et al suggested that social influence contributes to both
unpredictability and inequality of success, with follow-up experiments
confirming these initial
findings~\cite{salganik2009web,salganik2008leading,Muchnik2013,vandeRijt2014}
(see also \cite{hedstrom2006experimental} for a perspective on the
\musiclab{} paper).

In recent work \cite{PLOSONESI}, Abeliuk et al reconsidered the
\musiclab{} setting by exploiting the generative model that reproduces
the experimental data \cite{krumme2012quantifying}. They showed that
social influence always benefits market efficiency, if participants
are presented with the optimal ranking, i.e., the ranking that
maximizes the probability of a download at each step. Computational
experiments also showed that the optimal ranking significantly
outperforms the ranking in which products are displayed by decreasing
popularity (known as popularity ranking).

In this paper, we analyze the properties of a simple ranking policy
for trial-offer cultural markets: the {\em quality ranking} that
simply orders the products by quality. We show a number of positive
results about the quality ranking, including the fact that social
influence always benefits market efficiency when
it is used and that the benefits of social influence and position bias
\cite{Lerman2014} are cumulative. Most importantly, we show that,
under the quality ranking, the market almost surely converges to a
monopoly, assigning the entire market share to the product of highest
quality. As a corollary, this implies that the market is optimal
asymptotically. These results are robust and do not depend on specific
values for the appeal and quality of the products.  Moreover,
computational experiments using the generative model of the
\musiclab{} \cite{krumme2012quantifying} indicate a fast convergence
to the asymptotic predictions: The quality ranking quickly assigns the
largest proportion of the market share to the product of highest
quality and outperforms other ranking policies. Also, there is only
minor unpredictability in the market under the quality ranking policy.

Our results provide a sharp contrast with the conclusions in
\cite{salganik2006experimental}. Under the quality ranking policy, the
market is entirely predictable: It converges to ``blockbusters'' and
this convergence is due to social influence.  In other words, {\em it
  is not social influence that makes markets unpredictable:
  Unpredictability depends on how social influence is used.}

It is also important to emphasize that the benefits of popularity, the
social influence signal used in this paper, have been validated
experimentally. Indeed, two recent studies
\cite{engstrom2014demand,viglia2014please} were conducted to
understand the relative importance of the different social signals on
consumer behaviour. The first study \cite{engstrom2014demand} surveyed
the downloads of more than 500,000 from the Android marketplace Google
play, while the second study \cite{viglia2014please} conducted an
online experiment ($n=168$) where participants were asked to rank
hotels based on the number of reviews and the average rating. Both
studies arrived at the same conclusion, namely that the popularity
signal (i.e., social preference) has a much stronger impact than
the rating signal (i.e., a state preference). These two studies provide
evidence that quantitative models with popularity as the social signal
are of significant importance.

The relationships between quality, popularity, and appeal have also
been studied experimentally. Stoddard \cite{stoddard2015popularity}
studied the relationship between the intrinsic article quality and its
popularity in the social news sites Reddit and Hacker news. The author
proposed a Poisson regression model to estimate the demand for an
article based upon its quality, past views and age among others. The
results obtained after an estimation of each intrinsic article from
these social news site showed that the most popular articles are
typically the articles with the highest quality.  Another study of
social influence was carried out by \cite{tucker2011does} . The
authors conducted a field experiment which showed that popularity
information may benefit products with narrow appeal significantly more
than those with a broad appeal.

The rest of this paper is organized as follows. Section
\ref{section:market} introduces trial-offer markets and Section
\ref{section:rankings} surveys the ranking policies used in this
paper. Section \ref{section-properties} presents the main properties
of the quality ranking, including the benefits of position bias and
social influence and the asymptotic convergence.  Section
\ref{section:experiments} describes the computational experiments and
Section \ref{section:conclusion} presents the final discussion. All the
proofs are in the appendix.

\section{Trial-Offer Markets}
\label{section:market}

This paper considers a trial-offer market in which participants can
try a product before deciding to buy it. Such models are now
pervasive in online cultural markets (e.g., books and songs). The
market is composed of $n$ products and each product $i \in
\{1,\ldots,n\}$ is characterized by two values:
\begin{enumerate}
\item Its {\em appeal} $A_i$ which represents the inherent preference
  of trying product $i$;

\item Its {\em quality} $q_i$ which represents the conditional
  probability of purchasing product $i$ given that it was tried.
\end{enumerate}
Each market participant is presented with a product list $\pi$: she
then tries a product $s$ in $\pi$ and decides whether to purchase $s$
with a certain probability. The product list is a permutation of
$\{1,\ldots,n\}$ and each position $p$ in the list is characterized by
its {\em visibility} $v_p > 0$ which is the inherent probability of
trying a product in position $p$. Since the list $\pi$ is a bijection
from positions to products, its inverse is well-defined and is called
a ranking. Rankings are denoted by the letter $\sigma$ in the
following, $\pi_i$ denotes the product in position $i$ of the list
$\pi$, and $\sigma_i$ denotes the position of product $i$ in the
ranking $\sigma$. Hence $v_{\sigma_i}$ denotes the visibility of the
position of product $i$.

This paper examines a number of questions about these markets, including
\begin{itemize}
\item What is the best way to allocate the products to positions?

\item Is it beneficial to display a social signal, e.g., the number of
  past purchases, to customers?

\item Is the market predictable?
\end{itemize}
The primary objective is to maximize the market efficiency, i.e., the
expected number of purchases. Note also that the higher this objective
is, the lower the probability that consumers try a product but then
decide not to purchase it. Thus, if we interpret this last action as
an inefficiency, maximising the expected efficiency of the market also
minimizes unproductive trials.

\paragraph{Static Market} In a static market, the probability of
trying product $i$ given a list $\sigma$ is
\[
p_{i}(\sigma) =  \frac{v_{\sigma_i} A_i}{\sum_{j=1}^n v_{\sigma_j} A_j}
\]
and the static market optimization problem consists of finding a ranking
$\sigma$ maximizing the expected number of purchases, i.e.,

\begin{align}\label{p-ranking_formulae}
\max_{\sigma \in S_n} \sum_{i=1}^n p_{i}(\sigma) \ q_i 
\end{align}

\noindent
where $S_n$ represents the symmetry group over $\{1,\ldots,n\}$.

\noindent
Observe that consumer choice preferences for trying the products are
essentially modelled as a discrete choice model based on a multinomial
logit \cite{luce1959} in which product utilities are affected by their
position. Abeliuk et al. \cite{abeliuk2015_4OR} studied a
generalization of the maximization problem \eqref{p-ranking_formulae}
where it is not required to show all products to consumers.  The
authors interpreted the problem as a optimal assortment problem with
position-bias and they proved that it can be solved in polynomial
time.

\paragraph{Dynamic Market with Social Influence}

The paper also considers a dynamic market where the appeal evolves
over time according to a social influence signal. Given a social
signal $d=(d_1,\ldots,d_n)$, the appeal of product $i$ becomes $A_i +
d_i$ and the probability of trying product $i$ given a list $\sigma$
becomes
\[
p_{i}(\sigma,d) =  \frac{v_{\sigma_i} (A_i + d_i)}{\sum_{j=1}^n v_{\sigma_j} (A_j + d_j)}.
\]
The dynamic market uses the number of purchases $d_{i,t}$ of product
$i$ at a time $t$ as the social signal. Hence, in a dynamic market
with $T$ steps, the expected number of purchases is defined by the
following recurrence:
\begin{flalign*}
& u_t(d)  =  \max_{\sigma \in S_n} \sum_{i=1}^n P_{i}(\sigma,d) (q_i (1 + u_{t+1}(d[d_i \leftarrow d_i + 1])) + (1 - q_i) u_{t+1}(d)) &  (t \in 1..T) \\
& u_{T+1}(d) = 0 &
\end{flalign*}
where $d[i \leftarrow v]$ represents the vector $d$ where element $i$
has been replaced by $v$. The value $u_1(\langle 0,\ldots,0\rangle)$
denotes the optimal expected profit over the time horizon.

Observe that the probability of trying a product depends on its
position in the list, its appeal, and its number of purchases at time
$t$. As the market evolves over time, the number of purchases
dominates the appeal of the product and the trying probability of a
product becomes its market share. Note also that a dynamic market with
no social influence simply amounts to solving the static market
optimization problem repeatedly. This is called the {\em independent
  condition} in the rest of the paper.

In the following, without loss of generality, we assume that the
qualities and visibilities are non-increasing, i.e.,
\[
q_{1}\geq q_{2}\geq\cdots\geq q_{n}
\]
and
\[
v_{1}\geq v_{2}\geq\cdots\geq v_{n}.
\]
We also assume that the qualities and visibilities are known. In
practical situations, the product qualities are obviously
unknown. However, the quality of the products can be recovered
accurately and quickly, either before or during the market execution
\cite{PLOSONESI}. We use
\[
a_{i,t} = A_i + d_{i,t}
\]
to denote the appeal of product $i$ under social influence at step
$t$. When the step $t$ is not relevant, we omit it and use $a_i$
instead for simplicity. Finally, also for simplicity, we sometimes
omit the range of indices in aggregate operators when they range over
the products.

\section{Rankings Policies}
\label{section:rankings}

This paper focuses on the {\em quality ranking} which simply orders the
products by quality, assigning the product of highest quality to the
most visible position and so on. In other words, with the above
assumptions, a quality ranking $\sigma$ satisfies $\sigma_i = i \;\;
(1 \leq i \leq n)$.

The quality ranking contrasts with the {\em popularity ranking} which
was used in \cite{salganik2006experimental} to show the
unpredictability caused by social influence in cultural markets. At
iteration $k$, the popularity ranking orders the products by the
number of purchases $d_{i,k}$.  Note that purchases do not necessarily
reflect the inherent quality of the products, since they depend on how
many times the products were tried.

The {\em performance ranking} was proposed in \cite{PLOSONESI} to show
the benefits of social influence in cultural markets. The performance
ranking maximizes the expected number of purchases at each iteration,
exploiting all the available information globally, i.e., the appeal,
the visibility, the purchases, and the quality of the products. More
precisely, the performance ranking at step $k$ produces a list
$\sigma_k^*$ defined as
\[
\sigma_k^* = \argmax_{\sigma\in S_n} \sum_{i=1}^n p_{i}(\sigma,d_k)\cdot q_i
\]
where $d_k = (d_{1,k},\ldots,d_{n,k})$ is the social influence
signal at step $k$.  The performance ranking uses the probability
$p_{i,k}(\sigma)$ of trying products $i$ at iteration $k$ given
ranking $\sigma$, as well as the quality $q_i$ of product $i$. The
performance ranking can be computed in strongly polynomial time and
the resulting policy is scalable to large markets
\cite{PLOSONESI}.

In the rest of this paper, {\sc Q-rank}, {\sc D-rank}, and {\sc
  P-rank} denote the policies using the quality, popularity, and
performance rankings respectively. The policies are also annotated
with {\sc SI} or {\sc IN} to denote whether they are used under the
social influence or the independent condition. For instance, {\sc
  P-rank(SI)} denotes the policy that uses the performance ranking
under the social influence condition, while {\sc P-rank(IN)} denotes
the policy using the performance ranking under the independent
condition. We also use {\sc rand-rank} to denote the policy that
simply presents a random order at each period.

Under the independent condition, the optimization problem is the same
at each iteration as mentioned earlier. Since the performance ranking
maximizes the expected purchases at each iteration, it dominates all
other policies under the independent condition.

\begin{proposition}[Optimality of Performance Ranking] {\sc P-rank(IN)} is the optimal policy
  under the independent condition.
\end{proposition}

\noindent
The following example shows that the performance ranking is not optimal
in the dynamic market.

\begin{example}
\label{ex:quality}
Consider the following instance with 3 products:
\begin{itemize}
\item Visibilities: $v_1=0.7$, $v_2=0.2$, and $v_3=0.01$.
\item Qualities: $q_1=0.8$, $q_2=0.5$, and $q_3=0.1$.
\item Appeals: $A_1=0.01$, $A_2=0.1$, and $A_3=0.9$.
\end{itemize}
In the first step, the performance ranking is $\sigma^* =[2,1,3]$,
The expected number of purchases at time 1 given ranking $\sigma^*$ is
\[
\lambda^* = \sum_i \frac{v_{\sigma^*_i}A_{i}}{\sum_{j}v_{\sigma^*_j}A_{j}}q_i = 0.463.
\]
The probability that product i is purchased in time 1 is given by
\[
P^*_i= \frac{v_{\sigma^*_i}A_{i}}{\sum_{j}v_{\sigma^*_j}A_{j}}q_i
\]
which yields
\[
P^*_1 = 0.0198,\; P^*_2 = 0.432,\; P^*_3 = 0.0111.
\]
At time 2, there are four possible states, depending upon whether some
product was purchased at time 1. Let $s_0$ be the state where no
product was purchased and let $s_i \;(i\in \{1,2,3\})$ be the state
where product $i$ was purchased at time 0. The performance rankings
for each state are: $\sigma^*(s_1) =[1,2,3]$ and
$\sigma^*(s_0)=\sigma^*(s_2) = \sigma^*(s_3)=[2,1,3]$. The expected numbers
of purchases for each state are thus given by
\[
\lambda^*(s_0) =  0.463, \;\lambda^*(s_1) = 0.783, \; \lambda^*(s_2)= 0.496, \; \lambda^*(s_3)= 0.423.
\]
Therefore, the expected number of purchases of the performance ranking
for the first two steps is
\begin{equation*}\label{eq:two_period}
\lambda^* + \sum^3_{i=1} P^*_i \cdot  \lambda^*(s_i) + \left(1- \sum^3_{i=1} P^*_i   \right)\lambda^*(s_0) = 0.946,
\end{equation*}
where the first term is the expected purchases at time 1 and the
second and third terms are the expected number of purchases at time 2
given state $s_i$ times the probability of state $s_i$.

Consider now the quality ranking which corresponds to the permutation
$\sigma^q =[1,2,3]$ which is fixed for all possible states. The
expected number of purchases for ranking $\sigma^q$ in each of the
four states are
\[
\lambda^q(s_0)= 0.458, \; \lambda^q(s_1)= 0.783, \; \lambda^q(s_2)= 0.494, \;   \lambda^q(s_3)= 0.380.
\]
The probabilities that product $i$ is purchased at time 1 are given by
\[
P^q_1 = 0.156, \; P^q_2= 0.278, \; P^q_3= 0.025.
\]
As a result, the expected performance of the quality ranking in two
periods is $0.975$, outperforming the performance ranking by about 3\%
in this case.
\end{example}


\section{The Properties of the Quality Ranking}
\label{section-properties}

This section describes the main properties of the quality ranking.

\paragraph{The Benefits of  Position Bias}

We first show that position bias always increases the expected number of
purchases when quality ranking is used.

\begin{thm}
\label{thm:q-position-bias}
Position bias increases the expected number of purchases under the
quality-ranking policy, i.e., for all visibilities $v_i$, appeals
$a_i$, and qualities $q_i$ $(1 \leq i \leq n)$, we have
\[
\frac{\sum_{i}v_{i}a_{i}q_{i}}{\sum_{j}v_{j}a_{j}}\geq\frac{\sum_{i}a_{i}q_{i}}{\sum_{j}a_{j}}.
\]
\end{thm}
\proofatend
  Let $\lambda = \frac{\sum_{i}v_{i}a_{i}q_{i}}{\sum_{j}v_{j}a_{j}}$
  be the expected number of purchases for the quality ranking. We have
\[
\sum_{i}v_{i}a_{i}\left(q_{i}-\lambda\right)=0.
\]
Consider the index $k$ such that $\left(q_{k}-\lambda\right)\geq0$ and
$\left(q_{k+1}-\lambda\right)<0$. Since $v_{1} \geq \ldots \geq v_n$, we have
\[
\sum_{i=1}^{k}v_{k}a_{i}\left(q_{i}-\lambda\right)+\sum_{i=k+1}^{n}v_{k}a_{i}\left(q_{i}-\lambda\right) \leq \sum_{i}v_{i}a_{i}\left(q_{i}-\lambda\right) = 0
\]
and, since $v_k\geq 0$,
\[
\sum_{i=1}^{n}a_{i}\left(q_{i}-\lambda\right) \leq 0.
\]
It follows that
$
\lambda \geq \frac{\sum_{i=1}^{n}a_{i}q_{i}}{\sum_{i=1}^{n}a_{i}}.
$
\endproofatend

\paragraph{The Benefits of Social Influence}

An important question in cultural markets is whether the revelation of
past purchases to consumers improves market efficiency. This section
shows that, under the quality ranking policy, the expected number of
purchases in the studied trial-offer model increases when past
purchases are revealed.  This indicates that both social influence and
position bias improve the market efficiency and their benefits are
cumulative. The proof uses a lemma from \cite{PLOSONESI} specifying a
sufficient condition for a ranking to benefit from social influence.

\begin{thm}
The expected number of purchases is non-decreasing over time for the quality ranking under social influence.
\end{thm}
\proofatend Let 
\[
\mathbb{E}[D_{t}]=\frac{\sum_{i}v_{i}a_{i}q_{i}}{\sum_{i}v_{i}a_{i}}=\lambda.
\]
denote the expected number of purchases at time $t$. The expected
number of purchases in time $t+1$ conditional to time $t$ is
\[
\mathbb{E}[D_{t+1}]=\sum_{j}\left(\frac{v_{j}a_{j}q_{j}}{\sum v_{i}a_{i}}\cdot\frac{\sum_{i\not=j}v_{i}a_{i}q_{i}+v_{j}(a_{j}+1)q_{j}}{\sum_{i\not=j}v_{i}a_{i}+v_{j}(a_{j}+1)}\right)+\left(1-\frac{\sum_{i}v_{i}a_{i}q_{i}}{\sum_{i}v_{i}a_{i}}\right)\cdot\frac{\sum_{i}v_{i}a_{i}q_{i}}{\sum_{i}v_{i}a_{i}}
\]
\[
=\sum_{j}\left(\frac{v_{j}a_{j}q_{j}}{\sum v_{i}a_{i}}\cdot\frac{\sum_{i}v_{i}a_{i}q_{i}+v_{j}q_{j}}{\sum_{i}v_{i}a_{i}+v_{j}}\right)+\left(1-\frac{\sum_{j}v_{j}a_{j}q_{j}}{\sum_{i}v_{i}a_{i}}\right)\cdot\lambda.
\]
We need to prove that
\begin{equation}
\mathbb{E}[D_{t+1}] \geq \mathbb{E}[D_{t}],\label{eq:1}
\end{equation}
which amounts to showing that 
\[
\sum_{j}\left(\frac{v_{j}a_{j}q_{j}}{\sum v_{i}a_{i}}\cdot\frac{\sum_{i}v_{i}a_{i}q_{i}+v_{j}q_{j}}{\sum_{i}v_{i}a_{i}+v_{j}}\right)+\left(1-\frac{\sum_{j}v_{j}a_{j}q_{j}}{\sum_{i}v_{i}a_{i}}\right)\cdot\lambda \geq \lambda.
\]
which reduces to proving
\[
\frac{1}{\sum_{i}v_{i}a_{i}}\sum_{j}\left[\frac{v_{j}^2a_{j}q_{j}}{\sum_{i}v_{i}a_{i}+v_{j}}\left(q_{j}-\lambda\right)\right] \geq 0
\]
or, equivalently, 
\begin{equation}\label{eq:condition1}
\sum_{j}\left[\frac{v_{j}^2a_{j}q_{j}}{\sum_{i}v_{i}a_{i}+v_{j}}\left(q_{j}-\lambda \right)\right] \geq 0.
\end{equation}

\noindent
Let $k=\min \{ i \in N | (q_i-\lambda )\geq 0 \} $, i.e., the smallest
index $k\in N$, such that $q_k\geq \lambda$. We have that
\[
\sum_{j=1}^{n}\left[\frac{v_{j}^2a_{j}q_{j} \left(q_{j}-\lambda\right)}{\sum_{i=1}v_{i}a_{i}+v_{j}}\right]=
\sum_{j=1}^k \left[\frac{v_{j}q_{j}}{\sum_{i}v_{i}a_{i}+v_{j}} a_{j}v_{j}\left(q_{j}-\lambda\right)\right]+
\sum_{j=k+1}^n\left[\frac{v_{j}q_{j}\left(q_{j}-\lambda\right)}{\sum_{i}v_{i}a_{i}+v_{j}}a_{j}v_{j}\left(q_{j}-\lambda\right)\right].
\]
By definition of $k$, the terms in the summation on the left are
positive and the terms in the summation on the right are negative.
Moreover, for any $c>0$ and $v_i,v_j \geq0$, we have that
\[
\frac{v_{i}}{c+v_{i}}\geq\frac{v_{j}}{c+v_{j}}\Leftrightarrow(c+v_{j})v_{i}\geq(c+v_{i})v_{j}\Leftrightarrow cv_{i}\geq cv_{j}\Leftrightarrow v_{i}\geq v_{j}.
\]
Since $v_{1}\geq v_{2}\geq\ldots\geq v_{n}\geq 0$, 
 \begin{equation}\label{eq:property1}
\frac{v_{1}}{\sum_{i}v_{i}a_{i}+v_{1}}\geq\frac{v_{2}}{\sum_{i}v_{i}a_{i}+v_{2}}\geq\ldots\geq\frac{v_{n}}{\sum_{i}v_{i}a_{i}+v_{n}}.
\end{equation}
Moreover, since the quality ranking rankes the products by quality and
$q_{1}\geq q_{2}\geq\ldots\geq q_{n}\geq 0$, Equation
(\ref{eq:property1}) and the definition of $k$ implies that
\[
\forall i \leq k: \; \frac{v_i q_i}{\sum_{j}v_{j}a_{j}+v_{i}} \geq \frac{v_k q_k}{\sum_{j}v_{j}a_{j}+v_{k}},
\]
\[
\forall i > k: \; \frac{v_i q_i}{\sum_{j}v_{j}a_{j}+v_{i}} \leq \frac{v_k q_k}{\sum_{j}v_{j}a_{j}+v_{k}}.
\]
This observation, together with the fact that the left-hand
(resp. right-hand) terms are positive (resp. negative), produces a
lower bound to the right-hand side of Inequality
(\ref{eq:condition1}):
\[
\sum_{j=1}^k \left[\frac{v_{j}q_{j}}{\sum_{i}v_{i}a_{i}+v_{j}} a_{j}v_{j}\left(q_{j}-\lambda\right)\right]+
\sum_{j=k+1}^n\left[\frac{v_{j}q_{j}}{\sum_{i}v_{i}a_{i}+v_{j}}a_{j}v_{j}\left(q_{j}-\lambda\right)\right]
\]
\[
\geq\frac{v_k q_k}{\sum_{i}v_{i}a_{i}+v_k}\sum_{j=1}^k\left[a_{j}v_{j}\left(q_{j}-\lambda\right)\right]+
\frac{v_k q_k}{\sum_{i}v_{i}a_{i}+v_k}\sum_{j=k+1}^n\left[a_{j}v_{j}\left(q_{j}-\lambda\right)\right].
\]
Now, by definition of $lambda$, 
\[
\lambda = \frac{\sum_{i=1}^n v_{i}a_{i}q_{i}}{\sum_{i=1}^n  v_{i}a_{i}}
\Leftrightarrow 
\lambda \sum_{i=1}^n  v_{i}a_{i} = \sum_{i=1}^n  v_{i}a_{i}q_{i}
\Leftrightarrow 
\sum_{i=1}^n  v_{i}a_{i}(q_{i}-\lambda) = 0.
\]
which implies that 
\[
\frac{v_k q_k}{\sum_{i}v_{i}a_{i}+v_k}\sum_{j=1}^{n}\left[a_{j}v_{j}\left(q_{j}-\lambda\right)\right]=0
\]
concluding the proof.
\endproofatend

\paragraph{Asymptotic Behavior of the Quality Ranking}

We now show the key result of the paper: The trial-offer market
becomes a monopoly for the best product when the quality ranking is
used at each step. As a consequence, the quality ranking is optimal
asymptotically since the best product has the highest probability to
be purchased. The result also indicates that trial-offer markets are
predictable asymptotically.  To prove the result, we first
characterize the probability that the next purchase is product
$i$. Then we reduce a trial-offer market with two products to a
generalized P\'{o}lya scheme, a urn and ball model with
replacement. The result is then generalized to many products.

The first result characterizes the probability of the next purchase.
\begin{lem}
\label{p_i}
The probability $p_i$ that the next purchase (after any number of
steps) is product $i$ is
\begin{equation*}
p_i=\frac{v_i a_i q_i}{\sum\limits_{j=1}^n v_j a_j q_j}.
\end{equation*}
\end{lem}
\proofatend
The probability that product $i$ is purchased in the first step is given by
\begin{equation*}
p_i^{1st}= \frac{v_i a_i}{\sum\limits_{j=1}^n v_j a_j} q_i.
\end{equation*}
The probability that product $i$ is purchased in the second step and no product was purchased in the first step is given by
\begin{equation*}
  p_i^{2nd}=\left( \frac{\sum\limits_{j=1}^n v_j a_j(1-q_j)}{\sum\limits_{j=1}^nv_j a_j}\right)\frac{v_i a_i}{\sum\limits_{j=1}^nv_j a_j}q_i.
\end{equation*}
More generally, the probability that product $i$ is purchased in step
$m$ while no product was purchased in earlier steps is given by
\begin{equation*}
\label{noarreg}
p_i^{mth}=\left( \frac{\sum\limits_{j=1}^n v_j a_j (1-q_j)}{\sum\limits_{j=1}^n v_j a_j}\right)^m \frac{v_i a_i}{\sum\limits_{j=1}^nv_j a_j}q_i.
\end{equation*}
Defining $a=(\sum\limits_{j=1}^n v_j a_j q_j)/(\sum\limits_{j=1}^n v_j a_j)$, Equation \ref{noarreg} becomes
\begin{equation*}
p_i^{mth}=\bigg(1-a\bigg)^m  \frac{v_i a_i}{\sum\limits_{j=1}^nv_j a_j} q_i.
\end{equation*}
Hence the probability that the next purchased product is product $i$ is given by
\begin{equation*}
p_i=  \sum\limits_{m=0}^\infty\bigg(1-a\bigg)^m \frac{v_i a_i}{\sum\limits_{j=1}^nv_j a_j} q_i.
\end{equation*}
Since
\begin{equation*}
\sum\limits_{m=0}^\infty\bigg(1-a\bigg)^m=\frac{1}{a},
\end{equation*}
the probability that the next purchase is product $i$ is given by
\begin{equation*}
p_i=\frac{v_ia_iq_i}{\sum\limits_{j=1}^n v_ja_jq_j}.
\end{equation*}
\endproofatend

\noindent
Consider first a trial-offer
market with two products. Since the steps in which no product is
purchased can be ignored, by Lemma \ref{p_i}, we can use the following
variables $(1 \leq j \leq 2)$ to specify the market:
\begin{equation}
X_{j,t} \doteq a_{j,t} \hat{q}_j
\end{equation}
\begin{equation}
Z_{j,t} \doteq \frac{X_{j,t}}{T_t}
\end{equation}
\begin{equation}
T_{t} = T_{1,t} + T_{2,t}
\end{equation}
where $\hat{q}_j = v_j q_j$.

The trial-offer market can thus be modeled as generalized P\'{o}lya scheme
\cite{Renlund2010}, where $X_{j,t}$ represents the number of balls of
type $j$ at step $t$ and $Z_{j,t}$ is the probability that a ball of
type $j$ is drawn at step $t$. Since $X_{j,t+1} = X_{j,t} + \hat{q}_j$
if product $j$ is purchased, the generalized P\'{o}lya scheme add
$\hat{q}_j$ balls, each time product $j$ is purchased. As a result,
the P\'{o}lya scheme uses the replacement matrix
\begin{equation}
\label{replacement}
\Bigg(\begin{matrix}
  \hat{q_1} & 0 \\
  0 & \hat{q_2}
 \end{matrix}\Bigg).
\end{equation}

\noindent
This P\'{o}lya scheme is known to converge almost surely to a monopoly
whenever $\hat{q_1} \neq \hat{q_2}$ (see Theorem 6 in
\cite{Renlund2010}, whose proof uses stochastic approximation
techniques \cite{StochasticApproximation}).

\begin{lem}[Monopoly of a 2-Product Market]
\label{monopoly-2}
  Consider a Trial-Offer model with two products of quality $q_1$ and
  $q_2$ ($q_1 > q_2$). The market converges almost surely to a
  monopoly for product 1.
\end{lem}

\noindent
This lemma can be generalized when there are many
products but only two values of $\hat{q_i}$.

\begin{lem}
\label{monopoly-j}
  Consider a market with $m$ products such that $\hat{q}_1 = \ldots =
  \hat{q}_j = \hat{q}^+$ and $\hat{q}_{j+1}, \ldots, \hat{q}_n =
  \hat{q}^-$ ($1 \leq j \leq n$) with $\hat{q}^+ > \hat{q}^-$. Then, the market share of products
  $j+1,\ldots,n$ converges almost surely to zero.
\end{lem}
\proofatend
  The probability of purchasing one of the first $j$ (resp. last
  $n-j$) product next is
\begin{equation}
\frac{\sum\limits_{i=1}^j a_i \hat{q}^+}{\sum\limits_{i=1}^n a_i \hat{q}_i} (\mbox{\vspace{2cm} resp. } \frac{\sum\limits_{i=j+1}^n a_i \hat{q}^-}{\sum\limits_{i=1}^n a_i \hat{q}_i}).
\end{equation}
Hence, we can define a 2-product market in which product 1 has an
appeal $a^+$ and a quality $\hat{q}^+$ and product 2 has an appeal
$a^-$ and a quality $\hat{q}^-$, where
\begin{equation}
a^+ = \sum\limits_{i=1}^j a_i \mbox{\vspace{2cm} and } a^- = \sum\limits_{i=j+1}^n a_i.
\end{equation}
By Lemma \ref{monopoly-2}, this market converges almost surely to a
monopoly for product 1. Hence the market share for products
$j+1,\ldots,n$ converges almost surely to zero.
\endproofatend

\noindent
We are now in position to prove the main result of this section.

\begin{thm}[Monopoly of Trial-Offer Markets] Consider a trial-offer
markets where $q_1 > \ldots > q_n$. Then the market converges almost surely
to a monopoly for product 1.
\end{thm}
\proofatend
  Consider the generalized P\'{o}lya scheme $S$ associated with the market
  with $n$ types of balls. Define a new generalized P\'{o}lya scheme $S'$
  which is similar to $S$ except that $\hat{q}_1,\ldots,\hat{q}_{n-2}$
  are replaced by $\hat{q}_{n-1}$. At the limit, the fraction of balls
  of type $n$ in scheme $S'$ cannot be smaller than in scheme $S$.
  Consider now the market that corresponds to $S'$. By Lemma
  \ref{monopoly-j}, the market share of product $n$ converges almost
  surely to zero. The result follows from Lemma \ref{monopoly-2} after
  repeating the process for products $n-1,\ldots,3$.
\endproofatend

\noindent
This result has some key corollaries. In particular, the quality-ranking
is optimal asymptotically, since only the best-quality song is left.

\begin{corollary}
  The quality-ranking is asymptotically optimal in trial-offer
  markets.
\end{corollary}

\noindent
It is also important to point out that the market is completely
predictable at the limit.

For completeness, it is useful to consider the case where the products
have the same quality $q$ and there are no position bias. In this
case, the market share of each product converges almost surely to a
random variable following a beta-distribution. We prove this result
for the case of two products. By Lemma \ref{p_i}, the probability that
the next purchase is product $i$ becomes
\begin{equation}
p_i=\frac{a_i}{\sum\limits_{j=1}^n a_j}.
\end{equation}
and the replacement matrix becomes the diagonal matrix
$q\mathbb{I}^{n \times n}$. Now
\begin{equation}
Z_{i,k}\doteq\frac{a_{i,k}}{a_{1,k}+ a_{2,k}}.
\end{equation}
is a Martingale sequence. Indeed,
\begin{equation}
\begin{split}
\mathbb{E}_k[Z_{i,k+1}]=\frac{a_{i,k}+q}{a_{1,k}+a_{2,k}+q}Z_{i,k}+\frac{a_{i,k}}{a_{1,k}+a_{2,k}+q}(1- Z_{i,k})
\\
=\left(\frac{a_{i,k}+q}{a_{1,k}+a_{2,k}+q}\right)\frac{a_{1,k}}{a_{1,k}+ a_{2,k}}+\left(\frac{a_{i,k}}{a_{1,k}+a_{2,k}+q}\right)\frac{a_{2,k}}{a_{1,k}+ a_{2,k}}
\\
=\frac{a_{i,k}}{a_{1,k}+ a_{2,k}}\left(\frac{a_{1,k}+q}{a_{1,k}+a_{2,k}+q}+\frac{a_{2,k}}{a_{1,k}+a_{2,k}+q} \right)				\hspace{10mm}
\\
=\frac{a_{i,k}}{a_{1,k}+ a_{2,k}}=Z_{i,k} \hspace{40mm}
\end{split}
\end{equation}

\noindent
As a result, the market is equivalent to a classical
P\'{o}lya-Eggenberger urn model where the variable $Z_{i,k}$ converges
almost surely to a random variable obeying a beta distribution
(Theorem 3.2 in \cite{PolyaUrnModels}).

\begin{thm} Consider a market with two products of the same
  quality and equal visibilities. The market share of product $1$
  converges almost surely to a random variable obeying a beta
  distribution with parameters $A_1/q$ and $A_2/q$.
\end{thm}

\paragraph{Static Performance of Quality Ranking}

The previous section has shown that the quality-ranking is optimal
asymptotically in the dynamic market. We also know that the
performance ranking is optimal for the static market. This section
shows that the quality ranking has some performance guarantees even
for the static market.  In particular, it presents a bound on the
performance of the quality ranking which depends only the visibility
coefficients. The bound is also tight.

\begin{thm}[Static Performance Bound]
\label{thm:approx}
The quality ranking is an $\alpha$-approximation of the static market optimization problem,
where $\alpha=v_{1}/v_{n}$. Moreover, the $\alpha$-approximation of the quality ranking is tight.
\end{thm}
\proofatend
Let $\sigma^{*}$ be the optimal sorting and $\lambda^{*}$ its expected
number of purchases. We have
\[
\lambda^{*}=\frac{\sum_{i}v_{\sigma_{i}^{*}}a_{i}q_{i}}{\sum_{j}v_{\sigma_{j}^{*}}a_{j}}\leq\frac{\sum_{i}v_{1}a_{i}q_{i}}{\sum_{j}v_{n}a_{j}}=\alpha\frac{\sum_{i}a_{i}q_{i}}{\sum_{j}a_{j}}.
\]
Let $\lambda^{q}$ be the expected number of purchases for the quality ranking,
i.e.,
\[
\lambda^{q} = \frac{\sum_{i}v_{i}a_{i}q_{i}}{\sum_{j}v_{j}a_{j}}.
\]
By Theorem \ref{thm:q-position-bias},
\[
\lambda^{q}\geq\frac{\sum_{i}a_{i}q_{i}}{\sum_{j}a_{j}}.
\]
Combining both bounds yields $ \lambda^{*} \leq \alpha \lambda^{q}.  $

We now show that the approximation is tight. Consider 3 products with
qualities $q_{1}=1,q_{2}=\epsilon,q_{3}=0$ and appeals
$a_{1}=1,a_{2}=x,a_{3}=0$ and let the visibilities be
$v_{1}=1,v_{2}=1,v_{3} < 1$. The quality ranking is
$\sigma^{q}=(1,2,3)$ and the optimal performance ranking is
$\sigma^{*}=(1,3,2)$. The expected number of purchases for the quality
ranking is
\[
\frac{1+\epsilon x}{1+x}
\]
while it is
\[
\frac{1+\epsilon x\alpha}{1+\alpha x}.
\]
for the performance ranking.
When $\epsilon$ tends to zero, the ratio between the performance and
quality ranking becomes
\[
\lim_{\epsilon\rightarrow0}\frac{1+ v_3 \epsilon x }{1+v_3 x}\frac{1+x}{1+\epsilon x}=\frac{1+x}{1+v_3x}.
\]
Hence, when $x$ is large enough, the ratio is approximately $\alpha$, i.e., 
\[
\frac{1+x}{1+v_3 x}\approx\frac{1}{v_3}=\frac{v_{1}}{v_{3}}=\alpha.
\]
\endproofatend


\section{Computational Experiments}
\label{section:experiments}

The previous sections have derived theoretical results for the quality
ranking. In particular, they have shown that social influence and
position bias always increase the expected performance of the quality
ranking. These results hold for all possible values of the product
quality, appeal, and visibility inputs. In addition, the results have
provided strong static and asymptotic guarantees for the quality
ranking. This section aims at illustrating these results on settings
that model the \musiclab{} experiments discussed in
\cite{salganik2006experimental,krumme2012quantifying,PLOSONESI}.  As
mentioned in the introduction, the \musiclab{} is a trial-offer market
where participants can try a song and then decide to download it.  It
is important to emphasize that the generative model of the \musiclab{}
\cite{krumme2012quantifying} uses the same model for consumer choice
preferences.

\paragraph{The Experimental Setting}

\begin{figure}[t]
\begin{centering}
\includegraphics[width=0.4\linewidth]{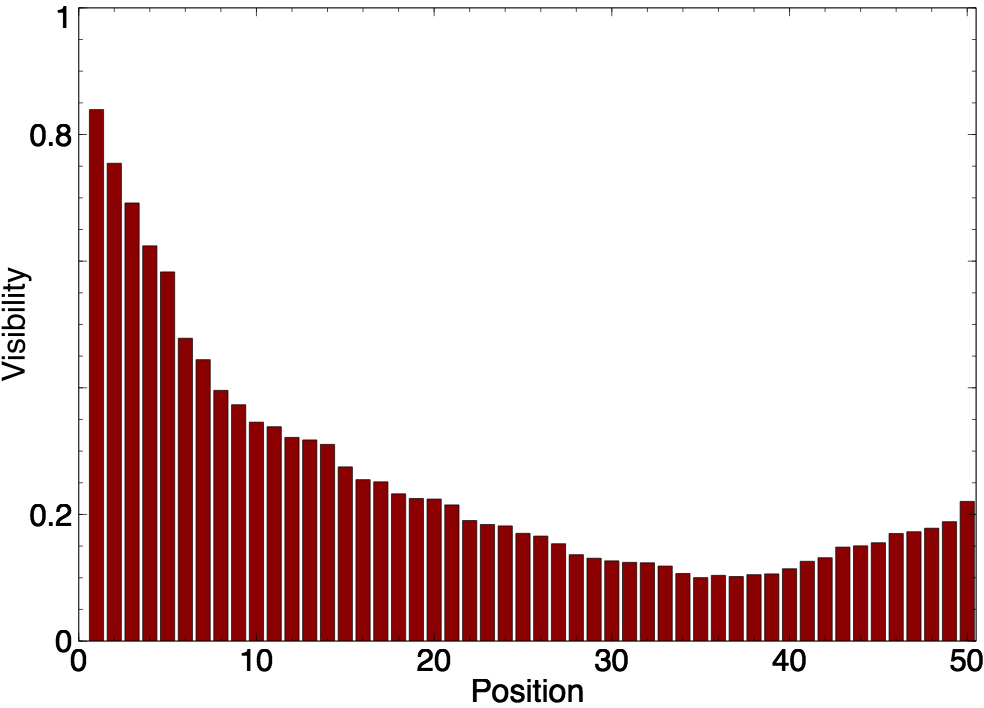}
\end{centering}
\centering{}
\caption{The visibility $v_p$ (y-axis) of position $p$ in the
  song list (x-axis) where $p=1$ is the top position and $p=50$ is
  the bottom position of the list which is displayed in a single column.}
\label{fig:visibility}
\end{figure}

\begin{figure}[t]
\begin{centering}
\includegraphics[width=0.75\linewidth]{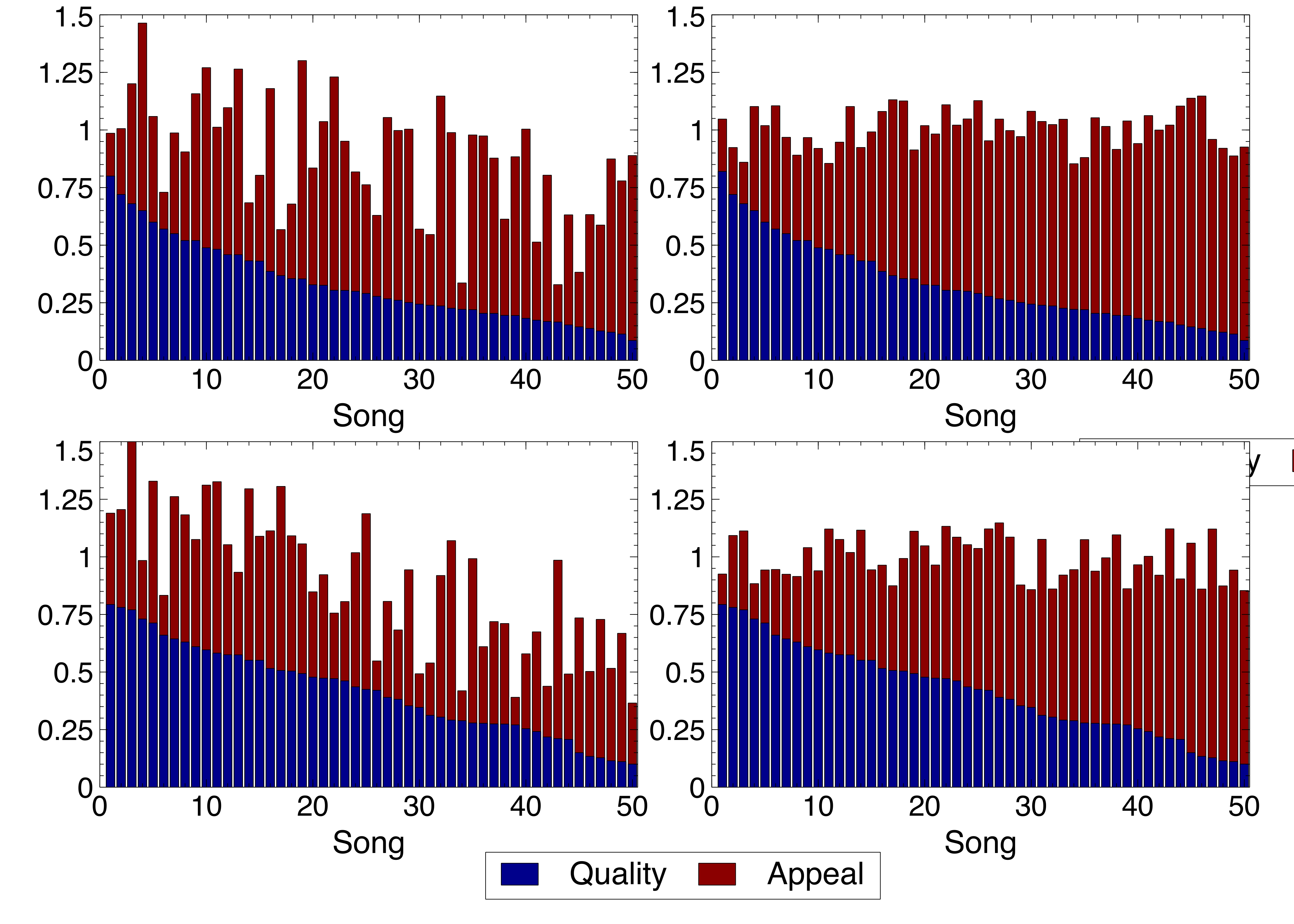}
\end{centering}
\centering{}
\caption{The Quality $q_i$ (blue) and Appeal $A_i$ (red) of song $i$
  in the four settings. In the first setting (top left), the quality
  and the appeal were chosen independently according to
  a Gaussian distribution. The second setting (top right) explores an
  extreme case where the appeal is negatively correlated with the
  quality used in setting 1. In the third setting (bottom left), the
  quality and the appeal were chosen independently
  according to a uniform distribution.The fourth setting (bottom
  right) explores an extreme case where the appeal is negatively
  correlated with the quality used in setting 3.}
\label{fig:qualityAppeal}
\end{figure}

The experimental setting uses an agent-based simulation to emulate the
\musiclab{}. Each simulation consists of $N$ iterations and, at each
iteration $t$,

\begin{enumerate}
\item the simulator randomly selects a song $i$ according to the
  probabilities $p_i(\sigma,d)$, where $\sigma$ is the ranking
  proposed by the policy under evaluation and $d$ is the social
  influence signal;

\item the simulator randomly determines, with probability $q_i$,
  whether selected song $i$ is downloaded; In the case of a download,
  the simulator increases the social influence signal for song $i$,
  i.e., $d_{i,t+1} = d_{i,t} + 1$. Otherwise, $d_{i,t+1} = d_{i,t}$.
\end{enumerate}

\noindent
Every $r$ iterations, a new list $\sigma$ is computed using one of the
ranking policies described above.  For instance, in the social
influence condition of the original \musiclab{} experiments, the
policy ranks the songs in decreasing order of download counts, i.e.,
the {\sc D-rank} policy. The parameter $r \geq 1$ is called the
refresh rate.

The experimental setting, which aims at being close to the \musiclab{}
experiments, considers 50 songs and simulations with 20,000
steps. The songs are displayed in a single column. The analysis in
\cite{krumme2012quantifying} indicated that participants are more
likely to try songs higher in the list. More precisely, the
visibility decreases with the list position, except for a slight
increase at the bottom positions. Figure \ref{fig:visibility} depicts
the visibility profile based on these guidelines, which is used in all
computational experiments. Under this setting, the (worst-case)
approximation factor of the quality ranking (see Theorem
\ref{thm:approx}) is $\alpha=v_{1}/v_{n}=0.8/0.2=4$.

The paper also uses four settings for the quality and appeal of each
product, which are depicted in Figure \ref{fig:qualityAppeal}:
\begin{enumerate}
\item In the first setting, the quality and the appeal were chosen
  independently according to a Gaussian distribution normalized to fit
  between $0$ and $1$.

\item The second setting explores an extreme case where the appeal is
  negatively correlated with quality. The quality of each product is
  the same as in the first setting but the appeal is chosen such that
  the sum of appeal and quality is exactly 1.

\item In the third setting, the quality and the appeal were chosen
  independently according to a uniform distribution.

\item The fourth setting also explores an extreme case where the
  appeal is negatively correlated with quality. The quality of each
  product is the same as in the third setting but the appeal is chosen
  such that the sum of appeal and quality is exactly 1.
\end{enumerate}
The experimental results were obtained by averaging the results of $W=400$ simulations.

\paragraph{Performance of the Market}

\begin{figure}[t]
\begin{centering}
\includegraphics[width=0.35\linewidth]{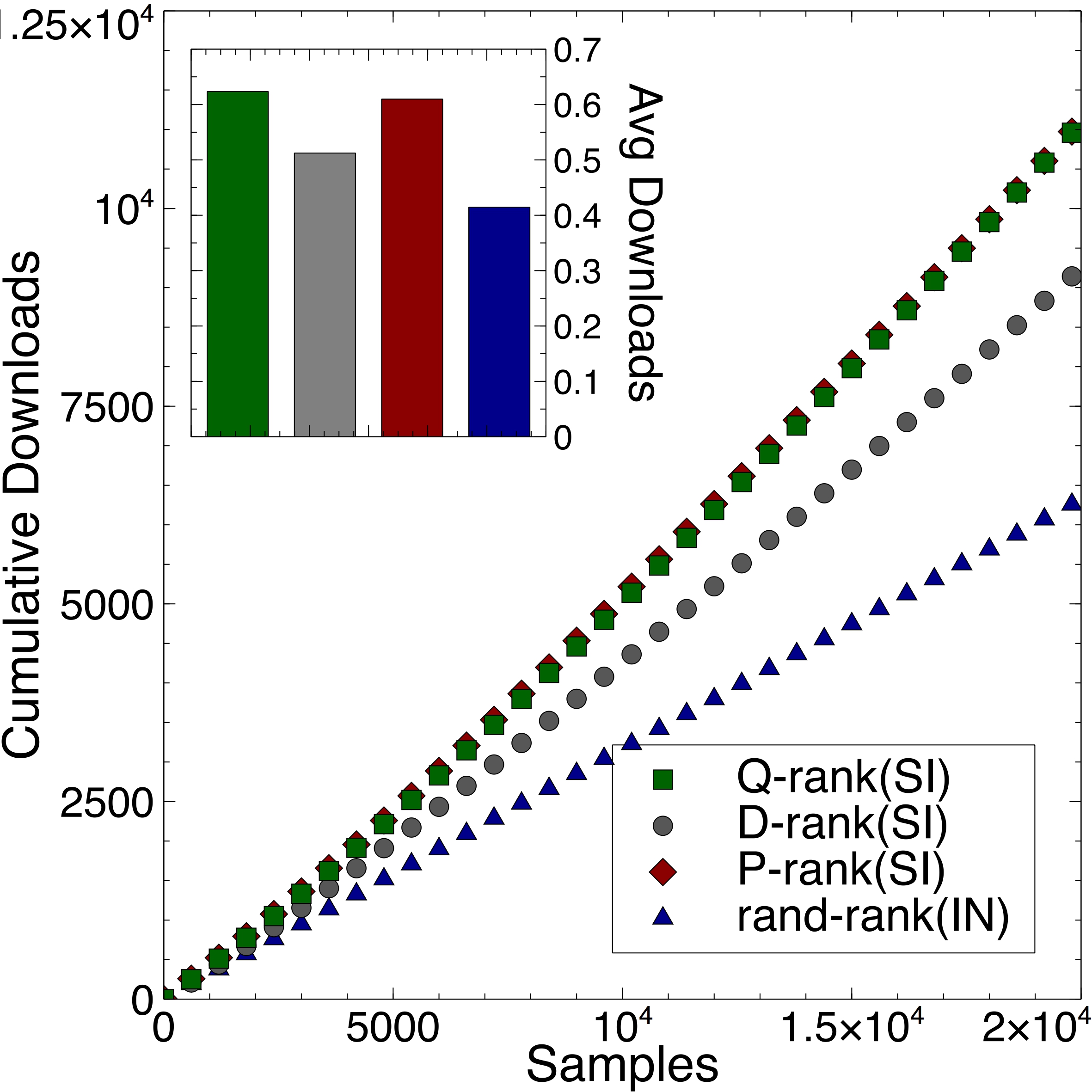}
\includegraphics[width=0.35\linewidth]{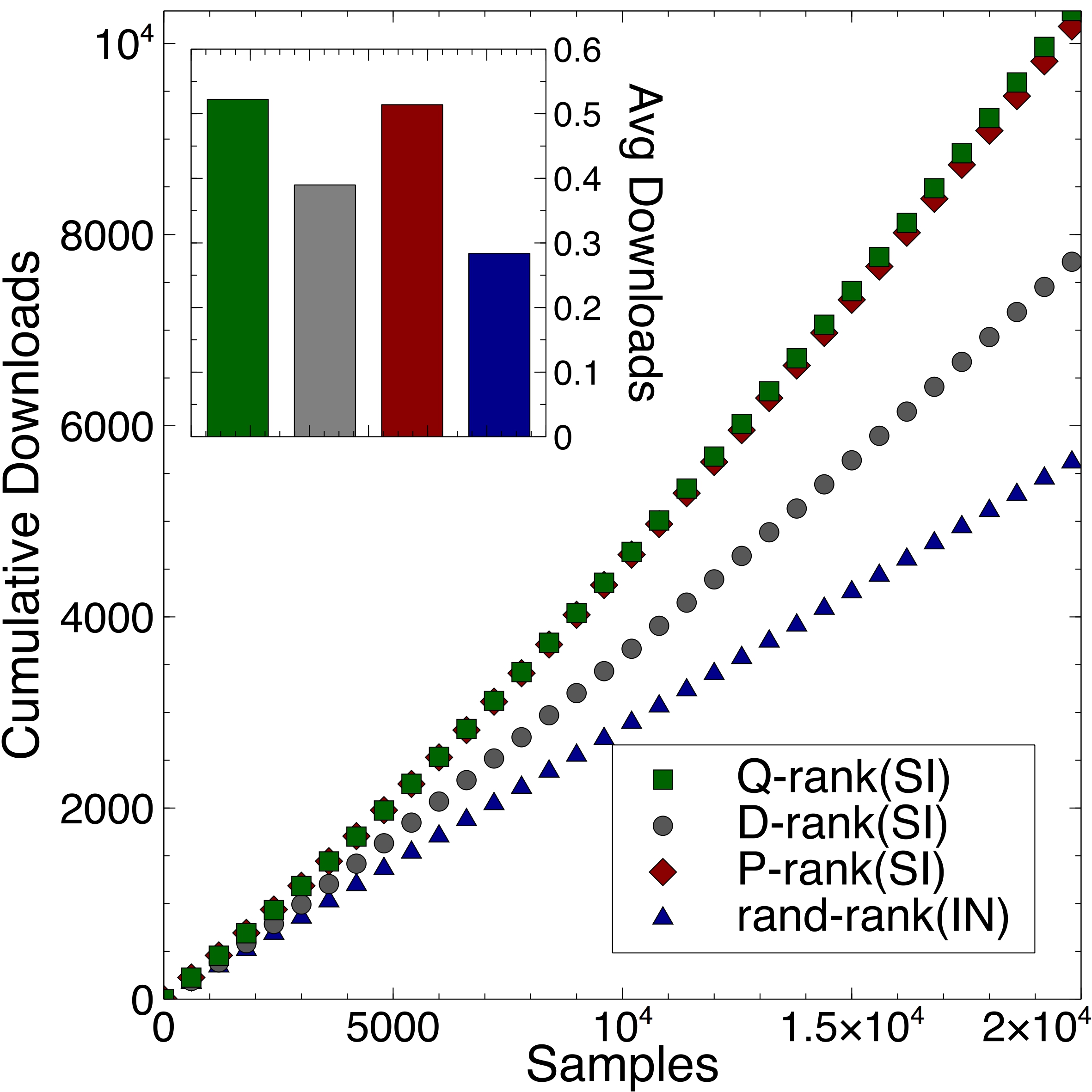}
\includegraphics[width=0.35\linewidth]{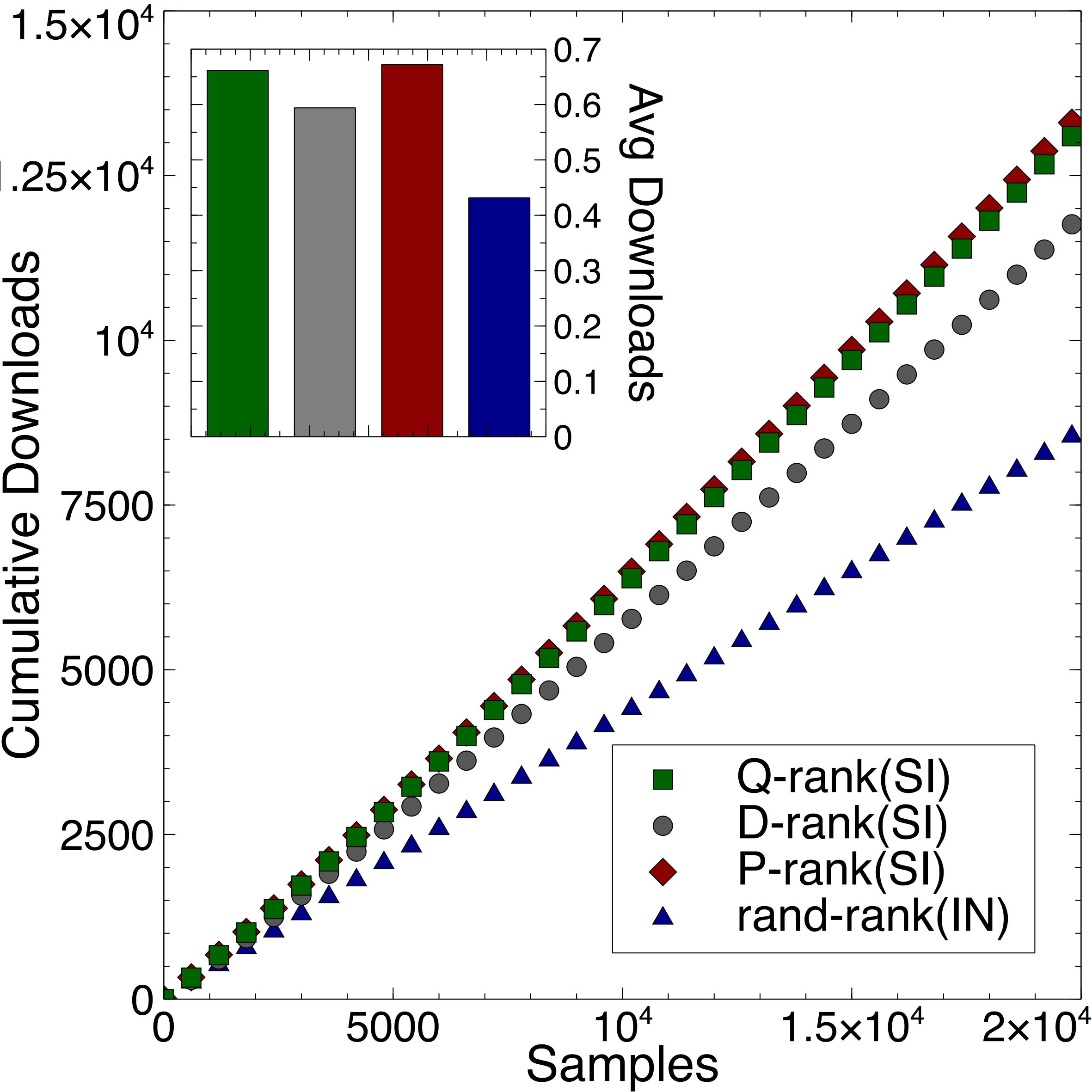}
\includegraphics[width=0.35\linewidth]{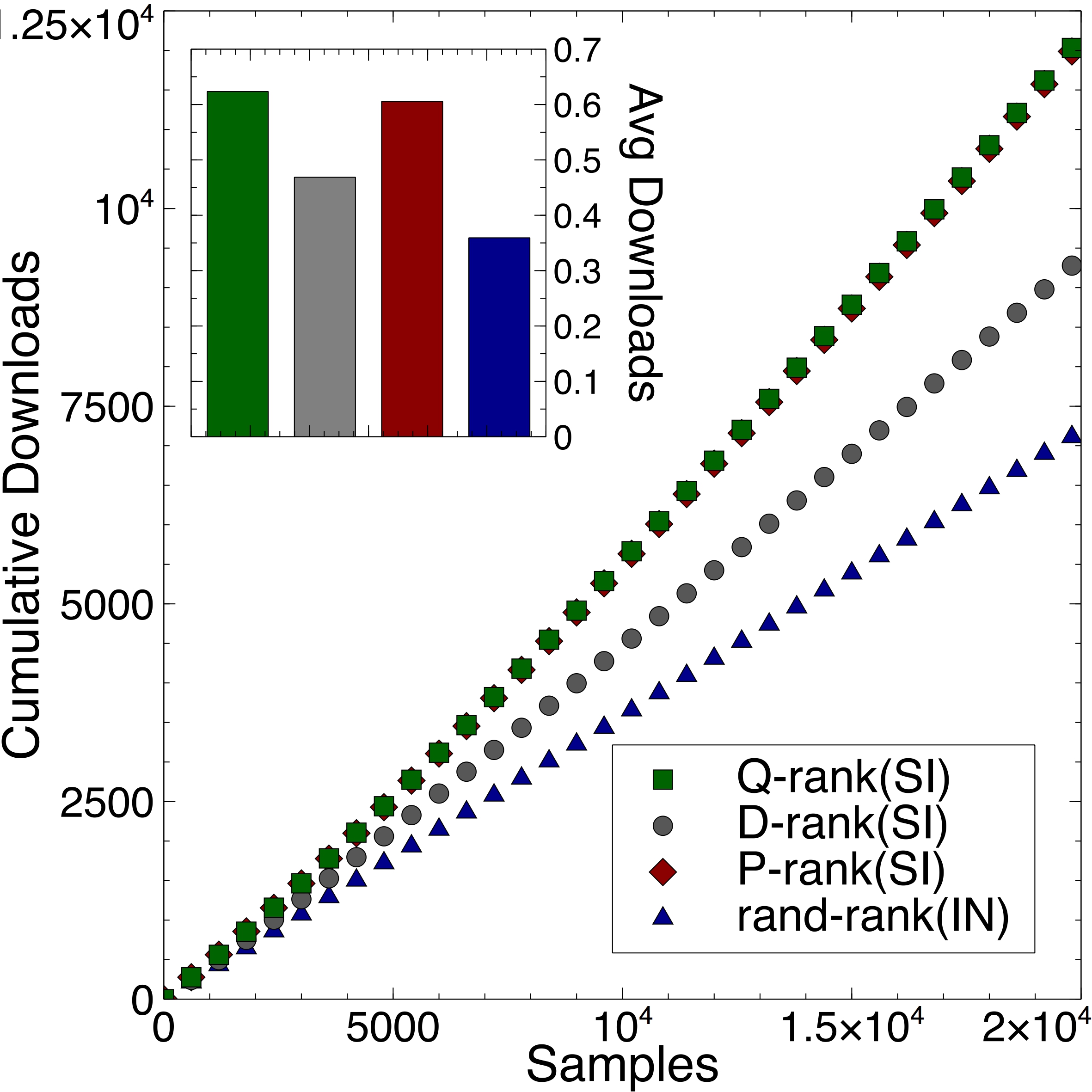}
\end{centering}
\centering{}
\caption{The Number of Downloads over Time for the Various
  Rankings. The x-axis represents the number of product trials and the
  y-axis represents the average number of downloads over all
  experiments. On the upper left corner of each graph, the bar plot
  depicts the average number of purchases per try for all
  rankings. The results for the four settings are shown in clockwise direction
  starting from the top-left figure.}
\label{fig:down}
\end{figure}

Figure \ref{fig:down} depicts computational results on the expected
number of purchases for the various rankings and settings. They
highlight two significant findings:
\begin{enumerate}
\item The quality ranking exhibits a similar performance to the
  performance ranking and provides substantial gains in expected
  downloads compared to the download (purchase) and random rankings.
  On settings with a negative correlations between appeal and quality,
  the quality ranking performs better than the performance ranking.

\item The benefits of social influence and position bias are
  complementary and cumulative as predicted by the theoretical
  results. Both are significant in terms of the expected performance
  of the market.
\end{enumerate}

\paragraph{Predictability of the Market}

\begin{figure}[t]
\begin{centering}
\includegraphics[width=0.70\linewidth]{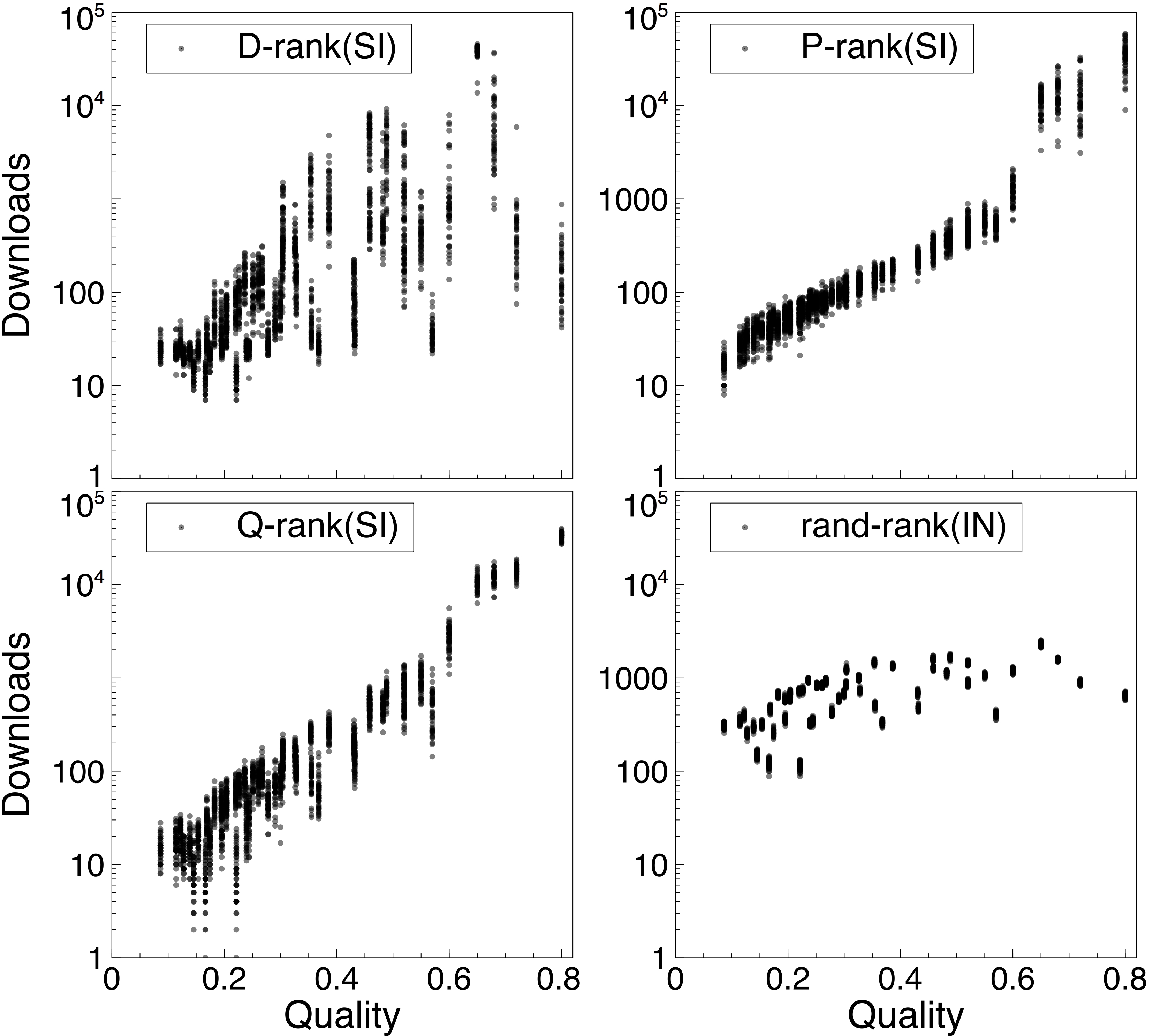}
\end{centering}
\centering{}
\caption{The Distribution of Downloads Versus Song Qualities
  (First Setting).  The songs on the x-axis are ranked by
  increasing quality from left to right. Each dot is the number of
  download of a product in one of the 400 experiments.}
\label{fig:q1}
\end{figure}

\begin{figure}[t]
\begin{centering}
\includegraphics[width=0.70\linewidth]{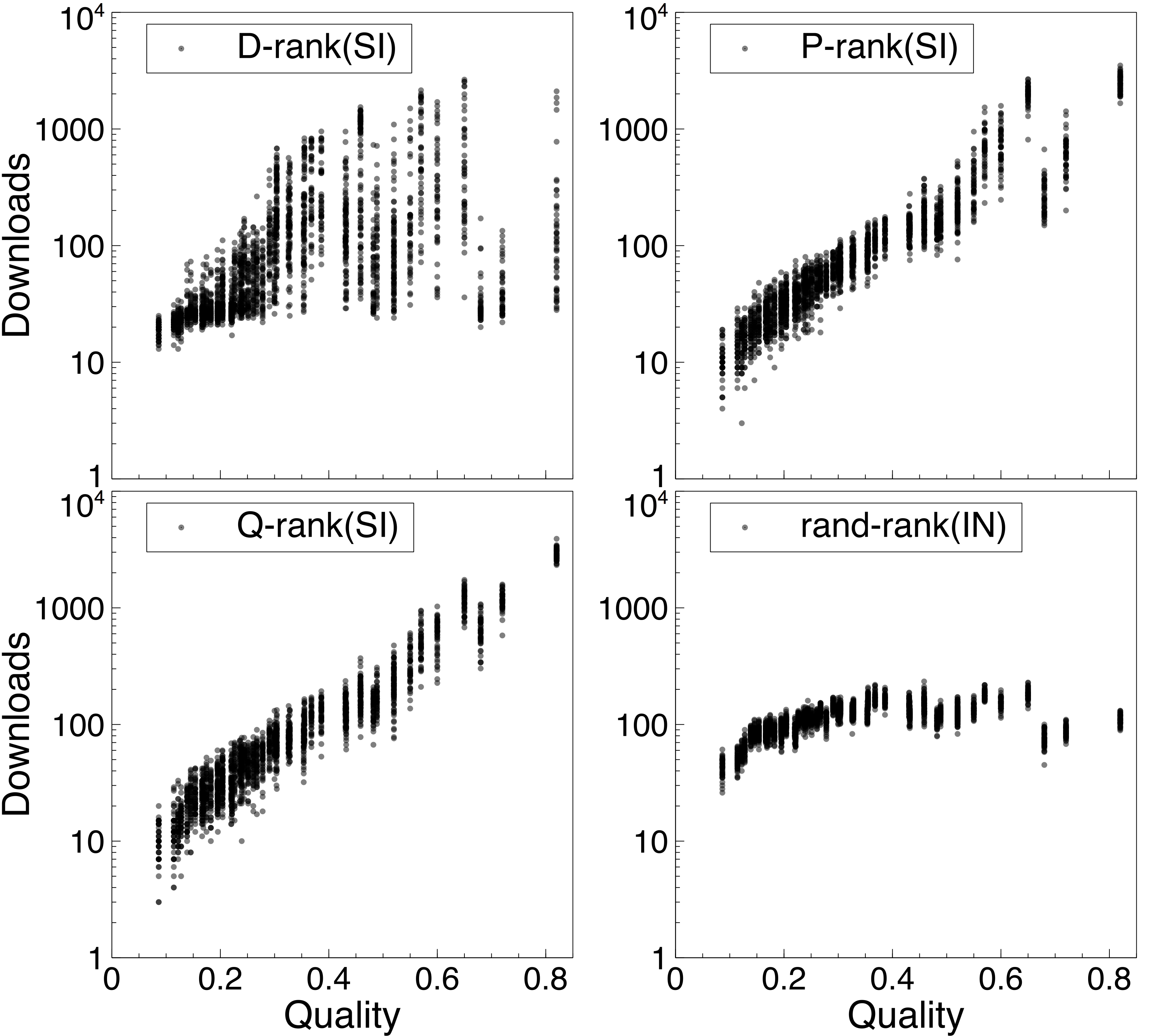}
\end{centering}
\centering{}
\caption{The Distribution of Download Versus Song Qualities
  (Second Setting).  The songs on the x-axis are ranked by
  increasing quality from left to right. Each dot is the number of
  downloads of a product in one of the 400 experiments.}
\label{fig:q2}
\end{figure}

Figures \ref{fig:q1} and \ref{fig:q2} depict computational results on
the predictability of the market under various ranking policies. The
figures plot the number of downloads of each song for the 400
experiments. In the plots, the songs are ranked by increasing quality
from left to right on the x-axis. Each dot in the plot shows the
number of downloads of a song in one of the 400 experiments. Figure
\ref{fig:q1} presents the results for the first setting where the
quality and the appeal were chosen independently according to a
Gaussian distribution. Figure \ref{fig:q2} presents the result for the
second setting where the appeal is negatively correlated with quality
and the quality of each song is chosen according to a Gaussian
distribution.

The computational results are compelling and confirms the theory.
Figure \ref{fig:q1} shows that the best song always receives the most
downloads in the quality ranking (with social influence) and that the
variance in its number of downloads across the experiments is very
small. The performance ranking (with social influence) also performs
well but the best song does not necessarily receive the most downloads
and the variance in its download is significantly larger. The download
ranking is highly unpredictable, while the random ranking is highly
predictable as one would expect. It is also interesting to note that
these observations continue to hold even when the appeal is negatively
correlated with quality, as Figure \ref{fig:q2} indicates. The
contrast between the popularity ranking used in
\cite{salganik2006experimental} and the quality ranking in this
setting is particularly striking.

In summary, the computational results confirm that, under the quality
ranking policy, the market is highly predictable. The song with the
highest quality emerges and the variance in its number of downloads is
very small.


\section{Discussion}
\label{section:conclusion}

This paper studied trial-offer cultural markets, which are ubiquitous
in our societies and involve products such as books, songs, videos,
clothes, and even newspaper articles. In these markets, participants
are presented with products in a certain ranking. They can then try
the products before deciding whether to purchase them.  Social
influence signals are widely used in such settings and help promote
popular products to maximize market efficiency. However, it has been
argued that social influence makes these markets unpredictable
\cite{salganik2006experimental}. As a result, social influence
is often presented in a negative light.

In this paper, we have reconsidered this conventional wisdom. We have
shown that, when products are presented to participants in a way that
reflects quality, the market is both efficient and predictable. In
particular, a quality ranking tends to a monopoly for the product of
highest quality, making the market both optimal and predictable
asymptotically. Moreover, we have also shown that both social
influence and position bias improve market efficiency. These results
are robust and do not depend on the particular values for the appeal
and quality of the products. In addition, computational experiments
using the generative model of the \musiclab{} show that there is a
fast convergence to our asymptotic theoretical results about the
performance and the predictability of the quality ranking. These
results are in sharp contrast with the popularity ranking studied in
\cite{salganik2006experimental}, where the products are ranked by the
strength of the social influence signals. In these conditions, the
market is indeed unpredictable and less efficient. There are some
important lessons to draw from these results.
\begin{itemize}
\item It appears that {\em unpredictability is not an inherent property of
  social influence}: Whether a market is predictable or not really
  depends on how social influence is used.

\item With the quality ranking, it is not hard to predict popularity
  as the market will converge to a monopoly. The computational
  experiments also show that ``blockbusters'' are quickly identified,
  even when the appeal is negatively correlated with quality.

\item Whenever the quality of the products can be approximated easily
  (e.g., by running some experiments with a small number of
  participants or learning it on the fly as suggested in
  \cite{PLOSONESI}), the quality ranking is the policy of choice: It
  is predictable and optimal asymptotically and is comparable, or
  outperforms, the performance ranking, i.e., greedy policy that
  continuously optimizes the ranking based on the appeal, the quality,
  and the social influence signal of each product.

\item With the quality ranking, a high-quality product overcomes a
  poor appeal but the opposite does not hold. A product of poor or
  average quality never become popular even with a great
  appeal. Obviously, aligning appeal and quality makes the market more
  efficient but it does not change the asymptotic outcome.
\end{itemize}

\noindent
It is also interesting to contrast our results with the study in
\cite{Ceyhan11} which uses a different model for consumer choice
preferences. In their model, the probability to purchase a
product is given by
\[
p_i(\phi) = \frac{e^{J \phi_i + q_i}}{\sum\limits_{j=1}^n e^{J \phi_j + q_j}}
\]
where $J$ is a constant and $\phi_i$ denotes the market share of
product $i$. With this choice model, with $J$ is large, monopoly
occurs and {\em ``eventually any of the products can get the largest
  market share, so the number of equilibria is $n$}''
\cite{Ceyhan11}. This comes from the fact that the social influence
signal is much stronger in this model, illustrating again that it is
how social influence is used, not social influence per se, which makes
markets unpredictable and less efficient. In other words, the
popularity ranking reinforces the social signal too much while, in the
model from \cite{Ceyhan11}, the social signal may be too
strong. Recall also that the choice model used in this paper was shown
to reproduce the original experiments of the \musiclab{}
\cite{krumme2012quantifying}.

Overall, these results paint a very different picture of the benefits
and weaknesses of social influence. Social influence does not make the
market unpredictable: When used with the quality ranking, it creates a
predictable market for our trial-offer model. Our current goal is to
obtain a broader understanding of social influence by generalizing
these results to different types of markets (e.g., markets with
heterogeneous customers) and different types of social signals.


\begin{thebibliography}{10}

\bibitem{abeliuk2015_4OR}
A.~{Abeliuk}, G.~{Berbeglia}, M.~{Cebrian}, and P.~{Van Hentenryck}.
\newblock {Optimizing Expected Profit in a Multinomial Logit Model with
  Position Bias and Social Influence}.
\newblock {\em ArXiv e-prints}, 2014.

\bibitem{PLOSONESI}
A.~Abeliuk, G.~Berbeglia, M.~Cebrian, and P.~Van~Hentenryck.
\newblock The benefits of social influence in optimized cultural markets.
\newblock {\em PLOS ONE}, 10(4), 2015.

\bibitem{Ceyhan11}
S.~Ceyhan, M.~Mousavi, and A.~Saberi.
\newblock {Social Influence and Evolution of Market Share}.
\newblock {\em Internet Mathematics}, 7(2):107--134, 2011.

\bibitem{engstrom2014demand}
Per Engstrom and Eskil Forsell.
\newblock Demand effects of consumers' stated and revealed preferences.
\newblock {\em Available at SSRN 2253859}, 2014.

\bibitem{hedstrom2006experimental}
Peter Hedstr{\"o}m.
\newblock Experimental macro sociology: Predicting the next best seller.
\newblock {\em Science}, 311(5762):786--787, 2006.

\bibitem{krumme2012quantifying}
Coco Krumme, Manuel Cebrian, Galen Pickard, and Sandy Pentland.
\newblock Quantifying social influence in an online cultural market.
\newblock {\em PloS one}, 7(5):e33785, 2012.

\bibitem{StochasticApproximation}
H.J. Kushner and G.G. Yin.
\newblock {\em Stochastic Approximation and Recursive Algorithms and
  Applications}.
\newblock Springer Verlag, 1997.

\bibitem{Lerman2014}
Kristina Lerman and Tad Hogg.
\newblock Leveraging position bias to improve peer recommendation.
\newblock {\em PLOS ONE}, 9(6):1--8, 2014.

\bibitem{luce1959}
D.~Luce.
\newblock {\em Individual Choice Behavior}.
\newblock John Wiley and Sons, 1965.

\bibitem{PolyaUrnModels}
Hosam Mahmound.
\newblock {\em Polya Urn Models}.
\newblock Chapman \& Hall/CRC Texts in Statistical Science, 2008.

\bibitem{Muchnik2013}
Lev Muchnik, Sinan Aral, and Sean~J. Taylor.
\newblock Social influence bias: A randomized experiment.
\newblock {\em Science}, 341(6146):647--651, 2013.

\bibitem{Renlund2010}
Henrik Renlund.
\newblock {Generalized Polya Urns Via Stochastic Approximation}.
\newblock {\em ArXiv e-prints 1002.3716}, February 2010.

\bibitem{salganik2006experimental}
Matthew~J Salganik, Peter~Sheridan Dodds, and Duncan~J Watts.
\newblock Experimental study of inequality and unpredictability in an
  artificial cultural market.
\newblock {\em Science}, 311(5762):854--856, 2006.

\bibitem{salganik2008leading}
Matthew~J Salganik and Duncan~J Watts.
\newblock Leading the herd astray: An experimental study of self-fulfilling
  prophecies in an artificial cultural market.
\newblock {\em Social Psychology Quarterly}, 71(4):338--355, 2008.

\bibitem{salganik2009web}
Matthew~J Salganik and Duncan~J Watts.
\newblock Web-based experiments for the study of collective social dynamics in
  cultural markets.
\newblock {\em Topics in Cognitive Science}, 1(3):439--468, 2009.

\bibitem{stoddard2015popularity}
Greg Stoddard.
\newblock Popularity and quality in social news aggregators: A study of reddit
  and hacker news.
\newblock {\em arXiv preprint arXiv:1501.07860}, 2015.

\bibitem{tucker2011does}
Catherine Tucker and Juanjuan Zhang.
\newblock How does popularity information affect choices? a field experiment.
\newblock {\em Management Science}, 57(5):828--842, 2011.

\bibitem{vandeRijt2014}
Arnout van~de Rijt, Soong~Moon Kang, Michael Restivo, and Akshay Patil.
\newblock Field experiments of success-breeds-success dynamics.
\newblock {\em Proceedings of the National Academy of Sciences},
  111(19):6934--6939, 2014.

\bibitem{viglia2014please}
Giampaolo Viglia, Roberto Furlan, and Antonio Ladr{\'o}n-de Guevara.
\newblock Please, talk about it! when hotel popularity boosts preferences.
\newblock {\em International Journal of Hospitality Management}, 42:155--164,
  2014.

\end{thebibliography}

\newpage

\section*{Proofs}

\printproofs

\end{document}